# Optimized structures for vibration attenuation and sound control in Nature: a review

Federico Bosia[1], Vinicius F. Dal Poggetto[2], Antonio S. Gliozzi[1], Gabriele Greco[2], Martin Lott[1], Marco Miniaci[3], Federica Ongaro[2], Miguel Onorato[4], Seyedeh F. Seyyedizadeh[1], Mauro Tortello[1], Nicola M. Pugno[2,5*]

[1] Department of Applied Science and Technology, Politecnico di Torino, Torino, 10129, Italy

[2] Laboratory for Bioinspired, Bionic, Nano, Meta Materials and Mechanics, Department of Civil, Environmental and Mechanical Engineering, University of Trento, Trento, 38123, Italy

[3] CNRS, Univ. Lille, Ecole Centrale, ISEN, Univ. Valenciennes, IEMN - UMR 8520, Lille, F-59000, France

[4] Department of Physics, University of Torino, Torino, 10125, Italy

[5] School of Engineering and Materials Science, Queen Mary University of London, London, E1 4NS, UK

* Corresponding author: nicola.pugno@unitn.it

**Summary**:

Nature has engineered complex designs to achieve advanced properties and functionalities through millions of years of evolution. Many organisms have adapted to their living environment producing extremely efficient materials and structures exhibiting optimized mechanical, thermal, optical properties, which current technology is often unable to reproduce. These properties are often achieved using hierarchical structures spanning macro, meso, micro, and nanoscales, widely observed in many natural materials like wood, bone, spider silk and sponges. Thus far, bioinspired approaches have been successful in identifying optimized structures in terms of quasi-static mechanical properties, such as strength, toughness, adhesion, but comparatively little work has been done as far as dynamic ones are concerned (e.g. vibration damping, noise insulation, sound amplification, etc.). In particular, relatively limited knowledge currently exists on how hierarchical structure can play a role in the optimization of natural structures, although concurrent length scales no doubt allow to address multiple frequency ranges. Here, we review the main work that has been done in the field of structural optimization for dynamic mechanical properties, highlighting some common traits and strategies in different biological systems. We also discuss the relevance to bioinspired materials, in particular in the field of phononic crystals and metamaterials, and the potential of exploiting natural designs for technological applications.

1. **Introduction**

It is well known that engineering materials such as metals or fibre-reinforced plastics are characterized by high stiffness at the expense of toughness. In particular, these materials do not efficiently dissipate energy via vibration damping. On the other hand, particularly compliant materials, such as rubbers and soft polymers, perform well as dampers, but lack in stiffness [1,2]. In this context, biological natural materials such as wood, bone, and seashells, to cite a few examples, represent excellent examples of composite materials possessing both high stiffness and high damping, and thus combine properties that are generally mutually exclusive. This exceptional behaviour derives from an evolutionary optimization process over millions of years, driven towards specific functionalities, where the natural rule of survival of the fittest has led to the continuous improvement of biological structure and organization. For instance, spider silk, bone, enamel, limpet teeth are examples of materials that combine high specific strength and stiffness with outstanding toughness and flaw resistance [3–8]. In these examples, a hierarchical architecture has often been proved to be the responsible for many energy dissipation and crack deflection mechanisms over various size scales, simultaneously contributing to exceptional toughness[2]. Given these numerous examples and the related interesting properties, the rich research field of biomimetics has emerged, with the aim of drawing inspiration from natural structures and implementing them in artificial systems, to bring progress to many technological domains.

Despite rapid progress in the field, studies in biomechanics and biomimetics linking material structure to function have mainly been limited to the quasistatic regime, while the dynamic properties of these materials have been somewhat less investigated, although notable examples of impact tolerance (e.g., the bombardier beetle's explosion chamber [9]) or vibration damping (e.g., the woodpecker skull [10]) have been studied. In fact, the first attempt to analyse biological vibration isolation mechanisms in the woodpecker date as far back as 1959, when Sielmann[11] found, through dissection and observation, that the cartilage in sutures in its skull have the effect of buffering and absorbing vibration[11]. Furthermore, recent studies have shown that structural hierarchy, which is typical of biological materials, can enhance the performance of artificial metamaterials [12,13]

As confirmed by these examples, it is reasonable to assume that biological structures whose main function is vibration and impact damping, sound filtering and focusing, transmission of vibrations, etc., have also been optimized through evolution, and that it is possible to look for inspiration in Nature for technological applications based on these properties. Based on this assumption, a growing interest in the superior vibration attenuation properties of biological systems has emerged, and nowadays, applications such as bio-inspired dampers are beginning to be used in the protection of precision equipment and the improvement of product comfort[14]. Motivated by this emerging field of research, we provide here a review of some of the main biological systems of interest for their dynamic properties, focusing on the role of structural architecture for the achievement of superior performance.

## 2. Impact resistant structures

### 2.1 Mantis shrimp

Probably the most well-known example of impact resistant structure in Nature is the stomatopod dactyl club. The mantis shrimp (*Odontodactylus scyllarus*) is a crustacean with a hammer-like club that can smash prey (mainly shells) with very high velocity impacts [15–17], reaching accelerations of up to 10000 *g*, and even generating cavitation in the water [18]. To sustain repeated impacts without failing, the claw requires extreme stiffness, toughness and impact damping, and has emerged as one of the main biological systems that epitomizes biological optimization for impact damage tolerance [19].

The exceptional impact tolerance is obtained thanks to the graded multiphase composition and structural organization of three different regions in the claw (Figure 1). The impact region, or striking surface, is dominated by oriented mineral crystals (hydroxyapatite), arranged so that they form pillars perpendicular to the striking surface. A second region called the "periodic region" backs up the impact zone and is mainly constituted by chitosan. This area, which lies just beneath the impact zone, is stacked at different (helicoidal) orientations, generating crack stopping and deviation. Thus, the structure consists of a multiphase composite of oriented stiff (crystalline hydroxyapatite) and soft (amorphous calcium phosphate and carbonate), with a highly expanded helicoidal organization of the fibrillar chitinous organic matrix, leading to effective damping of high-energy loading events [19,20]. The impact surface region of the dactyl club also exhibits a quasi-plastic contact response due to interfacial sliding and rotation of fluorapatite nanorods, leading to localized yielding and enhanced energy damping [21].

Interestingly, it has been found that the mantis shrimp also displays another highly efficient impact damping structure, since it has evolved a specialized shield in its tail segment called a telson, which absorbs the blows from other shrimps during ritualized fighting[22]. The telson is a multiscale structure with a concave macromorphology, ridges on the outside and a well-defined pitch-graded helicoidal fibrous micro-architecture on the inside, which also provides optimized damage tolerance [23,24].

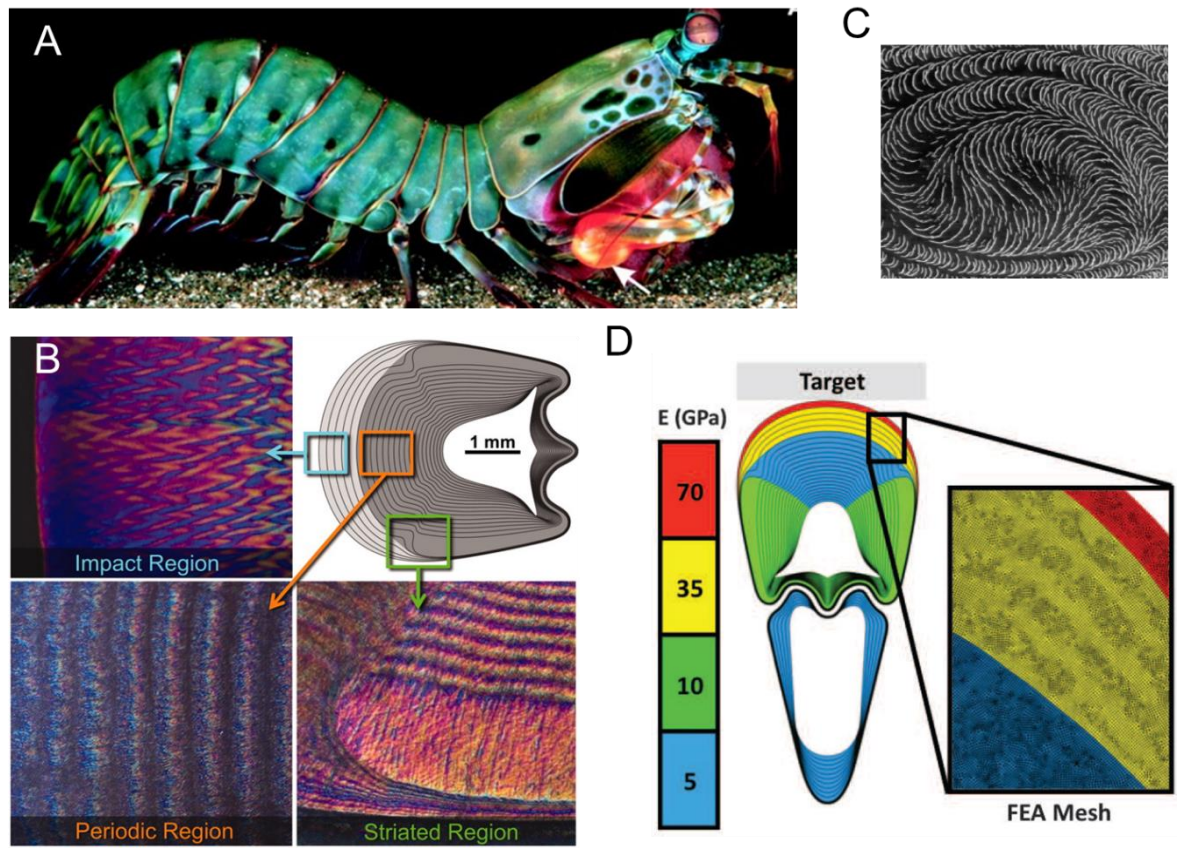

*Figure 1: A) Peacock mantis shrimp, with highlighted raptorial dactyl clubs to strike hard objects (adapted from [18]); B) Morphological features of the clubs, in cross-section view, divided in an impact region, a periodic region and a striated region; C) Scanning electron micrograph of the coronal cross-section, showing reinforcing fibre helicoidal arrangement; D) schematic of a Finite Element model accounting for graded material properties (adapted from [19]).*

### 2.2 Woodpecker skull

Another well-known example in Nature of a highly impact-resistant system is that of the woodpecker skull and beak, which repeatedly strikes wooden surfaces in trees at a frequency of about 20 Hz, a speed of up to 7 m/s, and can reach accelerations of the order of 1200 *g*, while avoiding brain injury [10,25]. This structure has been widely studied to draw inspiration for impact-attenuation and shock-absorbing applications and biomimetic isolation [14]. Limiting our observations to the head, and neglecting the body, feathers, and feet (which could also play a role), the woodpecker emerges as a very complex and rich system, from the mechanical and structural point of view at different spatial scales: macro-, micro- and nanoscale. The head is mainly formed by the beak, hyoid bone, skull, muscles, ligaments, and brain [26].

Several groups have investigated the mechanical behaviour of the woodpecker using finite element analysis [26–32]. Generally, the models are based on the images obtained by X-ray computed tomography (CT) scans. The stress distribution due to the impacts due to pecking is investigated. In some of these studies, the results are also compared with *in vivo* experiments, where the pecking force is measured by using force sensors and compared with that in other birds [27]. Zhu et al. [31] measured the Young's modulus of the skull, finding a periodic spatial variation, as reported in Figure 2A. Moreover, they performed a modal analysis on the skull by using a finite element model (Figure 2B), based on CT scan images, and determining the first ten natural frequencies, as shown in Figure 2C. The largest amplitude frequency components appear at 100 Hz and 8 kHz, which are well separated from the working frequency (around 20 Hz) and the natural frequencies (as derived in simulations), thus ensuring protection of the brain from injury.

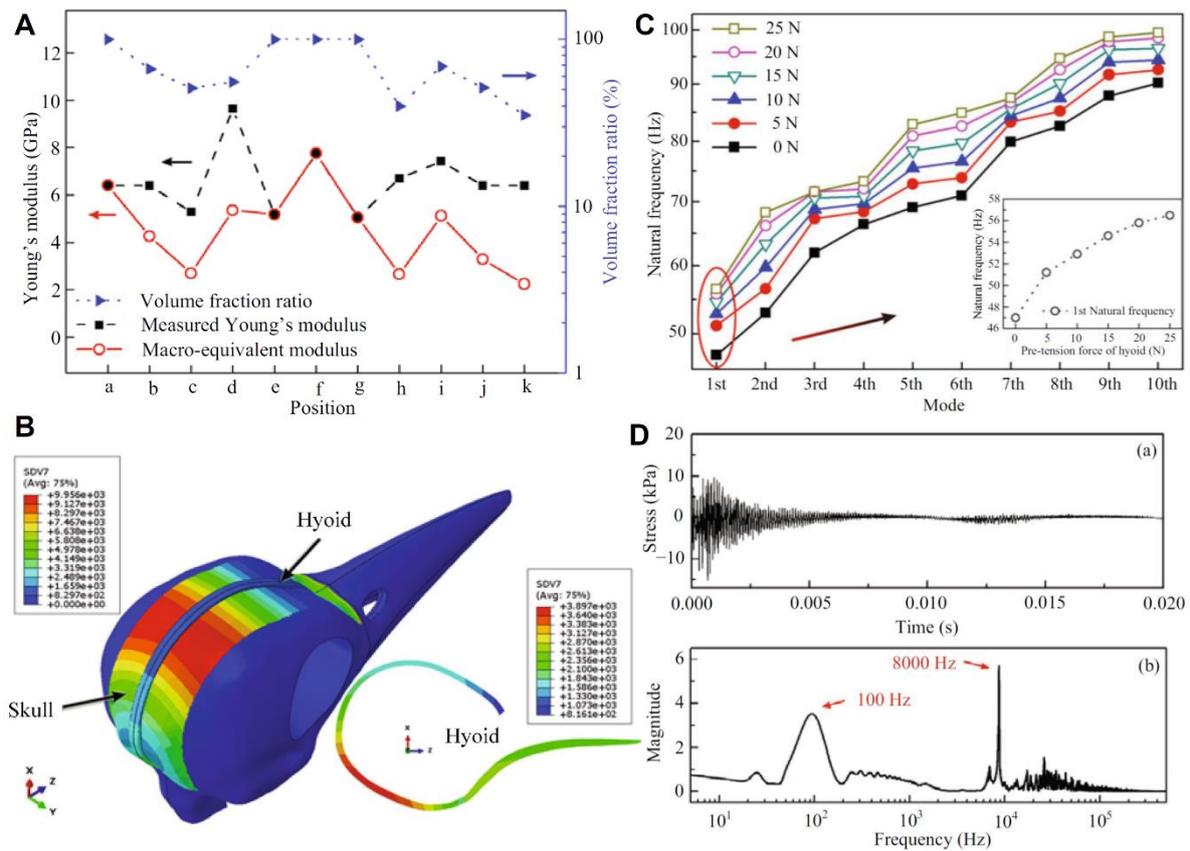

*Figure 2 : Vibration attenuation in the woodpecker skull (adapted from [31]). A): Volume fraction ratio of skull bone, local measured modulus, and macro-equivalent modulus around the skull. B): 3D finite-element model of the skull and hyoid bone, with spatial variation of the Young's modulus in the skull. C): first ten modes of the skull under a pre-tension on the hyoid in the range 0-25 N. D), upper panel: stress wave at a brain location under impact direction; lower panel: stress spectrum in the frequency domain obtained by FFT.*

Although the results from different groups are not always in agreement, most researchers conclude that the shape of the skull, its microstructure and material composition are all relevant for the exceptional impact-attenuation properties in woodpeckers [10]. In particular, a grading in the bone porosity and mechanical properties is particularly important in damping high frequency vibrations, which can be particularly harmful [33]. Many papers also point out the importance of the hyoid bone, very peculiar in woodpeckers, in the shock-absorption capability[34]. The hyoid is much longer than in other birds and wraps the skull until the eye sockets, forming a sort of safety belt around the skull. A specific study of the hyoid bone has been carried out by Jung et al. [34], who performed a macro- and micro-structural analysis of the hyoid apparatus and hyoid bones. The authors developed a 3D model of the hyoid which they showed it to be formed by four main parts connected by three joints. Interestingly, by performing nanoindentation measurements, they also showed that it features a stiffer, internal region surrounded by a softer, porous outer region which could play an important role in dissipating the energy during pecking. Another important issue is the relative contribution of the upper and lower beaks in the stress wave attenuation [27,35] which is most probably dissipated through the body [32].

Yoon and Park [10] showed that simple allometric scaling is not sufficient to explain the shock-absorbing properties of the woodpecker. Furthermore, they investigated its behaviour by using a lumped element model including masses, springs, and dampers, as shown in Figure 3A. Given the difficulty in modelling the complexity of the sponge-like bone within the skull with lumped elements, the authors characterized its behaviour by using an empirical method consisting of close-packed $SiO_2$ microglasses of different diameter (Figure 3B). The vibration transmissibility shows that the porous structure absorbs excitations with a higher frequency than a cut-off frequency which is determined by the diameter of the glass microspheres, as reported in Figure 3C.

Lee et al. [33] reported a detailed analysis on the mechanical properties of the beak, showing that the keratin scales are more elongated than in other birds and the waviness of the sutures between them is also higher than for other birds (1 for woodpecker, 0.3 for chicken and 0.05 for toucan), most probably to favour energy dissipation due to the impact. Raut et al. [36] designed flexural waveguides with a sinusoidal depth variation inspired by the suture geometry of the woodpecker beak which were tested by finite element analysis. The suture geometry helps reducing the group speeds of the elastic wave propagation whereas the presence of a viscoelastic material, as is the case for collagen in the beak sutures, significantly attenuates the wave amplitudes, suggesting a promising structure for applications in impact mitigation. Garland et al. [37] took inspiration from the same mechanism of the sliding keratin scales in the beak to design friction metamaterials for energy adsorption.

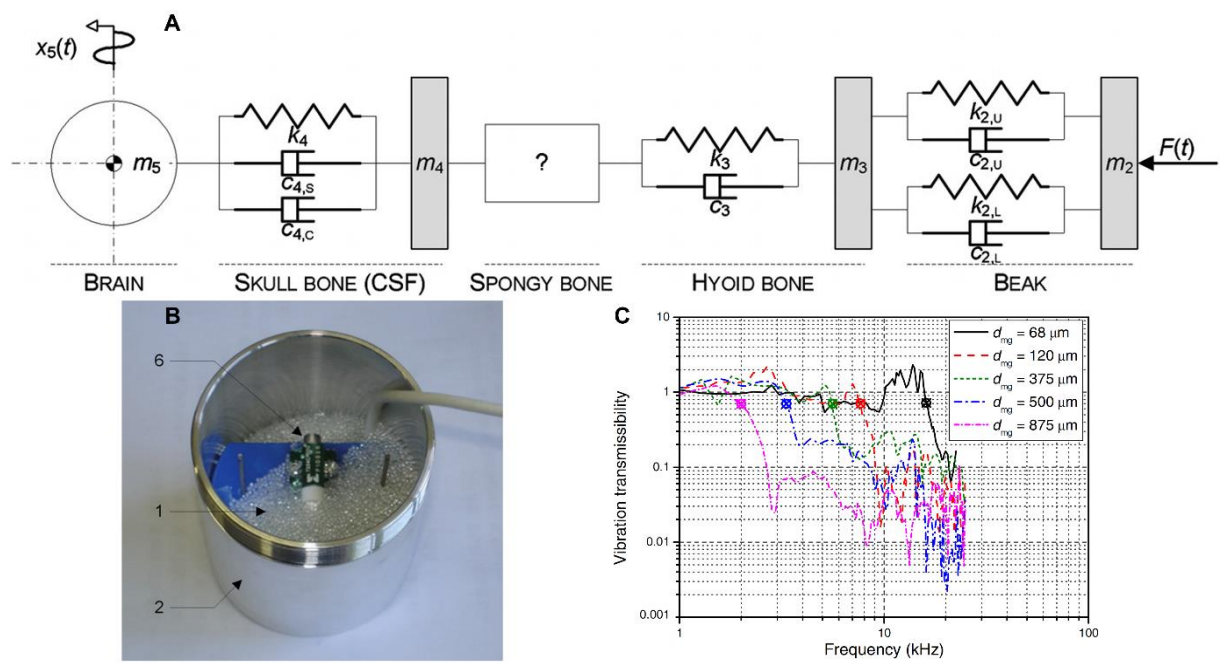

*Figure 3 : Modelling of vibration attenuation in the woodpecker skull (adapted from [10]). A): lumped-elements model of the head of a woodpecker. B): empirical model of the spongy bone by means of an aluminium enclosure filled with glass microspheres. C): vibration transmissibility as a function of frequency for different diameters of the SiO2 microspheres.*

### 2.3 Seashells

Seashells are rigid biological structures that are considered to be ideally designed for mechanical protection, and they are now viewed as a source of inspiration in biomimetics [38,39].

A seashell is essentially a hard ceramic layer that covers the delicate tissues of molluscs. Many gastropod and bivalve shells have two layers: a calcite outer layer and an iridescent nacre inner layer. Calcite is a prismatic ceramic material composed of strong yet brittle calcium carbonate ($CaCO_3$). Nacre, on the other hand, is a tough and pliable substance that deforms significantly before collapsing [40]. It is considered that a protective structure that combines a hard layer on the surface with a tougher, more ductile layer on the interior optimizes the impact damping properties [39–41]. When a seashell is exposed to a concentrated stress, such as a predator's bite, the hard ceramic covering resists penetration while the interior layer absorbs mechanical deformation energy. Overloading can cause the brittle calcite layer to fracture, causing cracks to spread into the soft tissue of the mollusc. Experiments have demonstrated that the thick nacreous layer can slow and eventually halt such fractures, delaying ultimate shell collapse. Although a significant amount of research has been performed on the structure and characteristics of nacre and calcite, there has been little research done on how these two materials interact in real shells. While there is evidence that nacre is tuned for toughness and energy absorption, little is known about how the shell structure fully utilizes its basic constituents, calcite, and nacre.

One method employed to analyse the geometry of the shell at the macroscale, while accounting for the micromechanics of the nacreous layer, is to adopt multiscale modelling and optimization [39]. Different failure modes are possible depending on the geometry of the shell. On the other hand, according to optimization procedures, when two failure modes in different layers coincide, the shell performs best in avoiding sharp penetration. To reduce stress concentrations, the shell construction fully leverages the material's capabilities and distributes stress over two different zones. Furthermore, instead of converging to a single point, all parameters converge to a restricted range inside the design space.

According to the experiments done on the two red abalone shells [39,42] the actual seashell arranges its microstructure design to fully utilize its materials and delay failure, a result that is also obtained through optimization. The crack propagates over the thickness of the shell in three different failure situations. Furthermore, the seashell, which is constructed of standard ceramic material, can resist up to 1900 N when loaded with a sharp indenter, which is an impressive load level given its size and structure.

### 2.4 Suture joints

Suture joints with different geometries are commonly found in biology from micro to macro length scales (Figure 4A) [43]. Examples include the carapace of the turtle [44,45], the woodpecker beak [33], the armoured carapace of the box fish [8,46], the cranium [47], the seedcoat of the *Portulaca oleracea* [48] and *Panicum miliaceum* [49], the diatom *Ellerbeckia arenaria* [50] and the ammonite fossil shells [51], among others.

In the aforementioned systems, the suture joint architecture, where different interdigitating stiff components, i.e., the teeth, are joined by a thin compliant seam, i.e., the interface layer, allows a high level of flexibility and is the key factor for the accomplishment of biological vital functions such as respiration, growth, locomotion and predatory protection [52–54]. Also, from a mechanical point of view, it has been demonstrated computationally and/or experimentally that this particular configuration allows an excellent balance of stiffness, strength, toughness, energy dissipation and a more efficient way to bear and transmit loads [54–58].

Several existing studies confirm this aspect. Among others [53,59], where, in the case of cranial sutures, it emerges that an increased level of interdigitation, found among different mammalian species, leads to an increase in the suture's bending strength and energy storage. Emblematic is the case of the leatherback sea turtle (Figure 4B), a unique specie of sea turtle having the capacity to dive to a depth of 1200 m [60]. This is due to the particular design of the turtle's carapace, where an assemblage of bony plates interconnected with collagen fibres in a suture-like arrangement is covered by a soft and stretchable skin. As reported in [60], the combination of these two elements provides a significant amount of flexibility under high hydrostatic pressure as well as exceptional mechanical functionality in terms of stiffness, strength and toughness, the collagenous interfaces being an efficient crack arrester. In addition, the study in [61] explained not only how the high sinuosity and complexity of the suture lines in ammonites (Figure 4C) are the result of an evolutionary response to the hydrostatic pressure, but also that the stress, displacements and deformations significantly decrease with the level of complexity. A similar result is also obtained in [62], which seeks to clarify the functional significance of the complex suture pattern in ammonites.

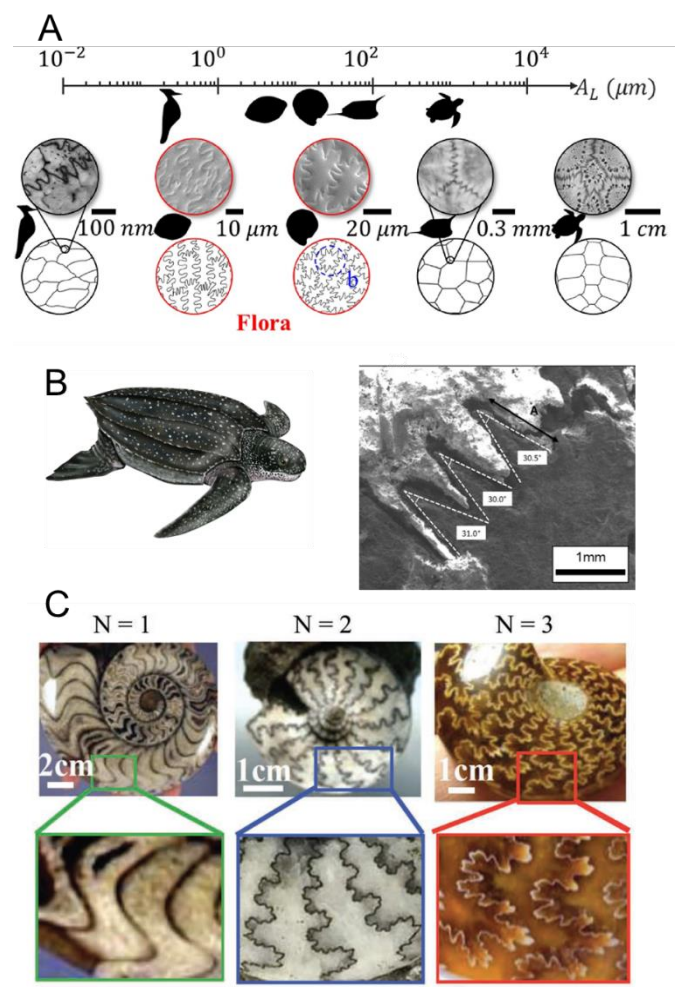

*Figure 4: Biological systems with suture tessellation: A) examples from Flora and Fauna (adapted from [58]); B) the leatherback turtle shell (adapted from [60]); C) Hierarchical sutures of increasing complexity found in ammonites (adapted from [63]).*

## 2.5 Bone

Bone has an extremely complex structure, encompassing seven levels of hierarchical organization, from nanocomposite mineralized collagen fibril upwards [64]. Based on this building block, varying mineral contents and microstructure allow to construct various types of tissue for different functionalities, e.g. withstanding tension, resisting impacts, supporting bending and compression. Human cortical bone consists of cylindrical Haversian canals, each surrounded by multilayers of bone lamellae ∼10 μm thick, which have a rotated plywood structure in which mineralized collagen fibrils (∼100 nm in diameter) rotate from the transverse direction to the longitudinal direction across five sublayers. The fibrils are cemented together by extrafibrillar minerals and noncollagenous proteins [6]. Hierarchical structure plays a fundamental role in bone's exceptional mechanical properties. Bone's trabecular structure and hierarchy is responsible for its unmatched tensile strength, anisotropy, self-healing and lightweight properties [64,65], but also dynamic properties like impact damping [66,67]. Bio-inspiration from bone structures has been exploited to seek enhanced static properties, strength and toughness [68], but relatively limited works have investigated it for dynamic applications. Ultrasonic wave measurements in bone to measure propagating velocity and attenuation have been performed for many years in various settings [69], including "wet" bone [70], showing that hydration is fundamental in defining dynamic properties. Studies in ultrasonics have typically focused on non-destructive evaluation of the bone structure [71,72]. It has also been shown that modal damping can be useful to detect bone integrity and osteoporosis [73], also supported by ultrasonic wave propagation simulations in cancellous bone [74]. Dynamic measurement methods assessing modal damping have also been used to validate bone models [75]. In terms of bioinspiration, the porous structure of trabecular (rod or truss-like structure) or velar bone (sail-like structure) is of particular interest, due to its lightweight and impact damping characteristics. Most of the work on such 3D frame structures [76] has addressed static properties [77]. However, recent articles have also addressed wave propagation [78,79] and impact loading [80]. Frame structures offer a convenient way to approximate trabecula using truss-like structures, inspired by the well-known Bravais lattices [81]. The implementation of such lattices could pave the way to a simplified model of the bone structure, where the joints can be collapsed to point-like connections and the number of degrees of freedom can be drastically reduced.

### 2.6 Attenuation of surface gravity waves by aquatic plants

If one considers damping of low frequency vibrations over long timescales, one can look to natural barriers that allow to prevent or delay coastal erosion, and the destruction of the corresponding habitats. One such example is the *Posidonia Oceanica*, a flowering aquatic plant endemic of the Mediterranean Sea, which aggregates in large meadows forming a Mediterranean habitat. This macrophyte has evolved by angiosperms typical of the intertidal zone and displays features similar to that of terrestrial plants: it has roots and very flexible thin leaves of about 1 mm thickness and 1 cm width, without significant shape variations along the leaf length. The anchoring to sandy bottoms is provided by the horizontal growth of the rhizomes, which also grow in vertical. The leave length varies throughout a year due to the seasonal cycle and the marine climatic conditions and can vary as much as 0.3 m in winter and 1 m in summer.

The effects of seagrasses on unidirectional flows are well studied at different scales in the field and in laboratory flumes and in numerical studies, while much less is known about the interaction between seagrass and waves. Wave attenuation due to *Posidonia* and flow conditions over and within vegetation fields have been investigated experimentally [82] and numerically [83]. It was found that the *Posidonia* is a good natural candidate for dissipating surface gravity waves in coastal regions. The study assessed quantitatively the physical value of the seagrass ecosystem restoration in this area, also opening new routes of action towards a resilient, efficient, and sustainable solution to coastal erosion. Other natural barriers to water wave propagation, other than vegetation such as the *Posidonia,* exist and are fundamental. For example, ice covering the surface of the ocean around Antarctica and the Artic Sea represents an important wave attenuation medium for slowing down the disintegration of the polar ice shelves. Quantitative measurements of such attenuation have been recently obtained through stereoscopic measurements [84].

### 2.7   Attenuation of surface seismic waves by trees

Further recent evidence of natural barriers for large scale vibrations is the attenuation of seismic surface waves achieved by trees [85]. The vibrations are transmitted to the trees through two coupling mechanisms, associated with two distinct vibrational modes. At high frequency (around 50 Hz), the longitudinal motion of the trees perpendicular to the soil surface is responsible for a high scattering effect on the surface wave and a hybridization to bulk shear waves. This means that the soil surface is mechanically blocked by the trees around those frequencies. In the low frequency range, below 1 Hz, the flexural motion of the trees induces different coupling effects on surface wave propagation. The flexural motion creates a bending moment at the soil/tree interface with can create long range coupling phenomena. Flexural resonances for the trees generally fall in the same frequency band as the micro-seismic noise produced by the ocean (between 0.3 and 0.8 Hz, detectable all over the world), which suggest a potential use of these frequencies to monitor the growth process of the trees and the evolution of their surrounding environment [86].

## 2.8 Conclusions on impact resistant structures

From the examples seen in the previous Sections, it emerges that impact-resistant biological structures have a number of common features. The first is related to a complex hierarchical architecture spanning from the nano- to the macro-scale, as in the case of the woodpecker skull or suture joints. Hierarchy, in particular, allows the system to be multifunctional and to accomplish both biological and mechanical functions in an optimized fashion. Additionally, in terms of dynamical behaviour, hierarchical structure allows to simultaneously address various size scales and therefore frequency ranges. The second characteristic is heterogeneity, enabling natural materials to combine the desirable properties of their building blocks, which are typically light, widely occurring materials: polymeric and ceramic for mineralized systems or crystalline and amorphous phases for non-mineralized ones. Heterogeneity allows Nature to create hierarchical composites that perform significantly better than the sum of their parts. Typically, the stiffer phase provides rigidity and strength while the soft phase increases ductility. This distinctive quality leads, for example, to the exceptional impact damping properties of the seashells described above. Another common trait of impact-resistant biological structures is porosity, which plays an important role in dissipating impact energy and, at the same time, allows to decrease the overall weight of the system. Finally, the occurrence of complex geometrical features is a characteristic commonly found in impact-resistant structures in biology. The high sinuosity of the suture lines in ammonites or the helicoidal organization of the mantis shrimp's dactyl clubs are examples of this, and direct evidence of their continually optimized nature, deriving from adaptation to the form that best achieves the required function.

### 3. Sensing and predation

### 3.1 Spider webs

Of all the natural structures that inspire and fascinate humankind, spider orb webs play a particularly central role and have been a source of interest and inspiration since ancient times. Spiders are able to make an extraordinary use of different types of silks to build webs which are the result of evolutionary adaptation and can deliver a compromise between many distinct requirements [87], such as enabling trapping and localizing prey, detecting the presence of potential predators, and serving as channels for intraspecific communication [88]. The variety of structures, compositions, and functions has led to the development of a large amount of literature on spider silks and webs [88–90] and their possible bio-inspired artificial counterparts [91,92].

The overall mechanical properties of spider orb webs emerge from the interaction between at least five types of silk [3,93], each with a distinct function in the web. The most important vibration-transmitting elements are made from the strong radial silk [94], which also absorbs the kinetic energy of prey [95,96] while sticky spiral threads, covered with glue, are used to provide adhesion to retain the prey [97,98]. Moreover, junctions within the webs can be composed of two different types of silk [93]: the strong and stiff piriform silk that provides strength to the anchorages [99,100] (Figure 5A-B), and the aggregate silk that minimizes damage after impacts [5,93] (Figure 5C). The mechanical synergy of such systems is therefore due to the mechanical response of the junctions [101], the constitutive laws of different types of silks, and the geometry of the webs [5]. The richness of these features, which are still the subject of many studies, have already inspired technologies with different goals in various scientific fields [102–104].

Spider orb webs are able to stop prey while minimizing the damage after impacts, thus maintaining their functionality [5], partially exploiting the coupling with aerodynamic damping that follows prey impacts [96]. This makes orb webs efficient structures for capturing fast-moving prey [105], whose location can then be detected due to the vibrational properties of the orb web. Efficiency in detecting prey by the spider is mediated by the transmission of signals in the webs, which needs to carry sufficient information for the prey to be located [106]. Using laser vibrometry, it has been demonstrated that the radial threads are less prone to attenuating the propagation of the vibrations compared to the spirals [87], due to their stiffer nature [107], allowing them to efficiently transmit the entire frequency range from 1 to 10 kHz.

Spiral threads can undergo several types of motion, including: (i) transverse (perpendicular to both the thread and the plane of the web) (ii) lateral (perpendicular to the thread but in the plane of the web), and (iii) longitudinal (along with the thread axis), thus yielding complex frequency response characteristics [108–110]. Distinct wave speeds are also associated with each type of vibration, i.e., transverse wave speed is determined by string tension and mass density, while longitudinal wave speed is linked to mass density and stiffness [111]. With the addition of more reinforcing threads due to the multiple lifeline addition by the spider, the orb webs appears to maintain signal transmission fidelity [112]. This provides further evidence of the impressive optimization achieved in these natural structures, which balance the trade-offs between structural and sensory functions.

The sonic properties of spider orb webs can also be significantly influenced by pre-stressing, as demonstrated in the study conducted by Mortimer et al. [113]. Wirth and Barth [114] have shown that silk thread pre-stress increases with the mass of the spider, considering both inter and intra-specific variations, and may be used to facilitate the sensing of smaller prey [115]. The pre-tension in webs can also be strongly influenced by large amplitude vibrations, as demonstrated by numerical analysis [116]. This dependence has been shown to be stronger if the structure is damaged, especially in the radial threads [117].

Investigations on the vibration transmission properties of silk have been conducted by accessing its high-rate stress-strain behaviour using ballistic impacts on Bombyx mori silk (which can be partially compared to spider silk) [118]. Some studies indicated that the capability of transmitting vibrations is relatively independent of environmental conditions such as humidity [119,120], but in general it is expected that they affect the silk Young's modulus and the pre-stress level on the fibres, and therefore the speed of sound (i.e., wave propagation speed) in the material [121–123]. This dependence is one of the reasons why the measurement of the speed of sound in silk has not produced homogeneous data [109,124,125], and could provide a possible degree of freedom for spiders in tuning the vibrational properties of their webs [113,124].

Spider orb webs have proven to be one of the most inspiring systems to design structures able to manipulate elastic waves. Although many types of webs can be extremely efficient in detecting and stopping prey [126], plane structures tend to be preferred when it comes to bio-inspired systems, due to their simplicity. Metamaterials can be designed exploiting the rich dynamic response and wave attenuation mechanism of orb webs [127], based on locally resonant mechanisms to achieve band gaps in desired frequency ranges [128], and further optimized to achieve advanced functionalities [129]. The possibility of designing low-frequency sound attenuators is also regarded as a common objective in metamaterials design, and spider web-inspired structures seem to be able to provide lightweight solutions to achieve this goal [130,131].

### 3.2 Spider sensing

Although many spiders have poor sight, remarkable sensors that make them capable of interacting with their surroundings have evolved [132], including hair-shaped air movement detectors, tactile sensors, and thousands of extremely efficient strain detectors (lyriform organs such as slit sensilla) capable of transducing mechanical loads into nervous signals embedded in their exoskeleton [133–135]. Air flow sensors, named trichobothria (Figure 5D), seem to be specifically designed to perceive small air fluctuations induced by flying prey, which are detectable at a distance of several centimetres [136]. Spiders can process these signals in milliseconds and jump to catch the prey using only the information about air flow [137]. Although this could be sufficient to guide the detection of the prey using trichobothria, it could be that different hair-like structures undergo viscosity-mediated coupling that affects the perception efficiency. Interestingly, in the range of biologically relevant frequencies (30−300 Hz), viscous coupling of such hair-like structures is very small [138]. It seems, in particular, that the distance at which two structures do not interact is about 20 to 50 hair diameters, which is commonly found in Nature [138,139]. Spiders are also equipped with strain sensors (lyriform organs), which are slits that occur isolated or in groups (Figure 5E) with a remarkable sensory threshold in terms of displacement (from 1.4 nm to 30 nm) and corresponding force stimulus (0.01 mN). Moreover, many of such organs have an exponential stiffening response to stimuli, which makes them suitable to detect a wide range of vibration amplitudes and frequencies. These organs act as filters with a typical high-pass behaviour [140] to screen the environmental noise found in Nature. Despite their remarkable capability in detecting vibration patterns (in frequencies between 0.1 Hz and several kHz), it is not yet clear how low-frequency signals are transmitted [141]. In any case, spider impact sensing on orb webs has been shown to be an intricate mechanism determined by both material properties and web structure [142].

The sensing capabilities of spiders have driven the design of bio-inspired solutions in terms of sensor technology. Materials scientists have designed bio-inspired hair sensors realized to work both in air [143,144] and water [145]. Furthermore, the lyriform organs have inspired crack-based strain sensors [146,147], eventually coupled with the mechanical robustness of spider silk [146]. Interestingly, these two types of structures (crack and hair sensors) may be combined in a multi-functional sensor. Results for such a spider-inspired ultrasensitive flexible vibration sensor demonstrated a sensitivity that outperforms many commercial counterparts [147].

Spider silk threads are also capable of detecting airflows by means of their fluctuation [148], providing an incredibly wide range of detectable frequencies, from 1 Hz to 50 kHz. Thus, by modifying these materials (e.g., making them conductive) it may be possible to produce devices able to expand the range of human hearing. It is clear, however, that many difficulties remain to be resolved to scale and fully optimize such bio-inspired solutions. Firstly, the reduction of the exposed surface can be large due to the need to integrate a sensor in the electronics. Secondly, wearing and application of the device could mechanically deteriorate its efficiency during its lifetime. Lastly, an engineering approach is in stark contrast with biological ones. In this context, a profound breakthrough is needed to achieve high efficiency in the self-assembly materials at the submicron scale.

### 3.3　Scorpion sensing

Scorpions are arachnids belonging to the *Subphylum Chelicerata* family of the arthropods (which includes spiders), which have evolved sensory mechanisms specially adapted to desertic environments [149]. Once structure-borne vibrations are produced in the ground, they propagate through bulk and surface waves: while the former propagate into the soil at large speeds and cannot be perceived by surface-dwelling animals, the latter can provide a useful information propagation channel for various species [150,151]. Sand offers an especially interesting medium in this regard, since its wave speed and damping are significantly lower than in other soils, favouring time-domain discrimination and processing [152]. Brownell [153] has shown that two types of mechanoreceptors can be observed in the *Paruroctonus Mesaensis* desert scorpion species: (i) sensory hairs on the tarsus, which sense compressional waves, and (ii) mechanoreceptors located at the slit sensilla, which sense surface waves, thus serving as the basis for the scorpion's perception of the target direction, performing a role of mechano-transduction similar to that observed in spiders [154]. Thus, these structures appear to be those responsible for vibration sensing in scorpions, even though some controversy exists regarding the use of other scorpion appendages for the same purpose [155]. Brownell and Farley have shown that this scorpion species can discriminate the vibration source direction by resolving the time difference in the activation of the slit sensilla mechano-receptors even for time intervals as small as 0.2 ms [156]. The same authors have also shown that for short distances (down to 15 cm), scorpions can discriminate not only direction but also distance and vibration signal intensities, which are means to distinguish between potential prey from potential predators [157]. Such underlying phenomena have been used to construct a computer theory that simulates prey-localizing behaviour in scorpions [158], further motivating the development of artificial mechanisms based on this approach. Microstructural investigations as the ones performed by Wang et al. [159] have demonstrated that the slit sensilla owe their micro-vibration sensing properties to their tessellated crack-shaped slits microstructure [160], further indicating that this type of microstructure can serve as a bioinspiration for the design of new mechano-sensing devices [146,161].

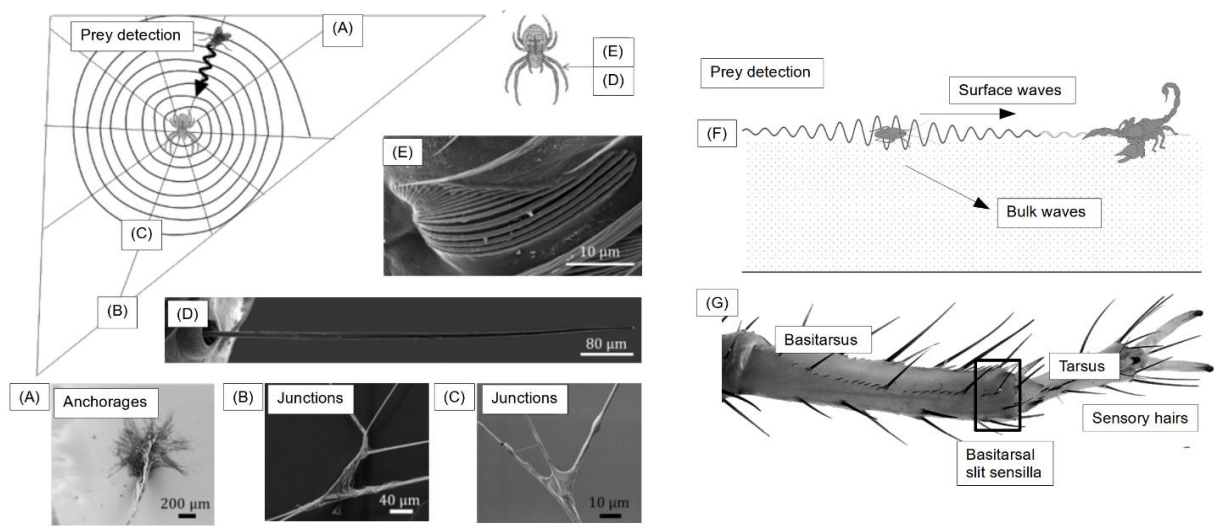

*Figure 5: Prey sensing similarities in spiders and scorpions. A) web structure: a typical orb web of a spider Nuctenea umbratica. The web is built by means of junctions between threads and surfaces, B) junctions between radial threads, C) and junctions between radial and spiral threads. A flying prey can be eventually detected by air flow sensors, D) the tricobothria. If the*

*prey impacts the web, the vibrational signal will be transmitted mainly by radial threads and be perceived by E) lyriform organs of the spider. Figure adapted from [93,132]. F) schematic of scorpion prey detection using surface waves; G) sensory hairs and mechanoreceptors located at the slit sensilla sense surface waves. Adapted from [159,162]*

### 3.4 Control of ground-borne sound by mammals

Vibration control in mammals is not restricted to air-borne signals. Many use impacts by "drumming" parts of their bodies to generate vibrations that propagate in the soil. For example, foot-drumming patterns may be found in rabbits or elephant to communicate with other individuals. Unlike in the Cochlea, where the signal is split, and thus analysed, according to its frequency content, foot-drumming is based on the generation and analysis of complex transient vibrational patterns. One of the first species identified to use foot-drumming to communicate is the blind rat [163]. More recent studies have identified the social and environmental monitoring purposes associated with this communication channel in elephants [164–166].

### 3.5 Anti-predatory structures and strategies

It is thought that the origin of many distinctive morphological and/or behavioural traits of living organisms is related to the selective pressure exerted by predators [167,168]. Generally, various defensive strategies can be adopted by organisms to reduce the probability of being attacked or, if attacked, to increase the chances of survival. The first consists in avoiding detection (i.e., *crypsis*), through camouflage, masquerade, *apostatic* selection, subterranean lifestyle or nocturnality, and deterring predators from attacking (i.e., *aposematism*) by advertising the presence of strong defences or by signalling their unpalatability by means of warning coloration, sounds or odours [169]. The second are based on overpowering, outrunning and diverting the assailants' strikes by creating sensory illusions to manipulate the predator's perception [170–172].

Despite being extremely fascinating from an engineering point of view, the effectiveness of the first type of defensive strategies is restricted mainly to visual phenomena and none of them work on non-visually oriented predators. However, although rare, several acoustic based deflection strategies exist in Nature. Most of them are related to one of the most famous examples of non-visually oriented predators, i.e., echolocating bats (Figure 6A) that rely on echoes from their sonar cries to determine the position, size and shape of moving objects in order to avoid obstacles and intercept prey in the environment [168,173–175].

The first strategy to avoid detection by bats can be seen in some species of earless moth that, as a result of millions of years of evolution, developed a passive acoustic camouflage relying on a particular configuration of both the thorax and the wings. In particular, differently from the other species of moth which evolved ears to detect the ultrasonic frequencies of approaching bats or produce, when under attack, ultrasound clicks to startle bats and alert them to the moth's toxicity [176–178], the wings of earless moths are covered with an intricate layer of scales (Figure 6B) that serve as acoustic camouflage against bat echolocation [177,179]. According to [177], each leaf-like shaped scale shows a hierarchical design, from the micro-to the nanoscale, consisting, at the larger scale, of two highly perforated laminae made of longitudinal ridges of nanometer size connected by a network of trabeculae pillars. This configuration leads to a highly porous structure which is able, because of the large proportion of interstitial honeycomb-like hollows, to absorb the ultrasound frequencies emitted by bats and thus reduce the amount of sound reflected back as echoes [180]. As a result, the moth partially disappears from the bat's biosonar and the distance at which the bat can detect the moth is reduced by 5-6% [179], representing a significant survival advantage. In addition, by exploring the vibrational behaviour of a wing of a *Brunoa alcinoe* moth, researchers discovered that each scale not only behaves like a resonant ultra-sound absorber having the first three resonances in the typical echolocation frequency range of bats [177], but also that each one has a different morphology and resonates at a particular frequency, creating a synergistically broadband absorption [180]. As reported in [180], it can be thus said that the complex pattern of scales on moth wings exhibit the key features of a technological acoustic metamaterial.

Another example of an acoustic-based strategy to confuse predators is the long hindwing tail (Figure 6C) commonly found on luna moths (*Actias luna*). Such tail presents a twist toward the end and this distinguishing feature, as suggested in [181], is the key for how the tail creates a sort of acoustic camouflage against echolocating bats. The tail, in particular, because of its length and twisted morphology, in reflecting the bat's sonar calls produces two types of echoic sensory illusions [181]. The first consists in deflecting the bat's attacks from the vital parts of the body, i.e., head and thorax, to this inessential appendage. By using high-speed infrared videography to analyse the bat-moth interactions, according to the authors, in over half of the interactions, bats directed the attack at the moth's tail as the latter created an alternative target distracting from the principal one, i.e., the moth's body. Also, by comparing moths with the tail and moths with the tail ablated, it emerged a survival advantage of about 47%.

The second sensory illusion provided by the twisted tail consists in inducing a misleading echoic target localization that confuses the hunting bats [170,181]. As reported in [181], the origin of this effect is the twist located at the end of the tail that creates a sequence of surfaces having different orientations so that, independently of the inclination of both the incident sound waves and the fluttering moth, the tail is able to return an echo, complicating and spatially spreading the overall echoic response of the moth. In addition, the analysis of the overall acoustic return generated by the wings, body and tail of a Luna moth, revealed an additional survival contribution of the twisted tail, consisting in a shift of the echoic target centre, i.e., the centre of the echo profile used by the bat to estimate the prey location, away from the moth's thorax [181].

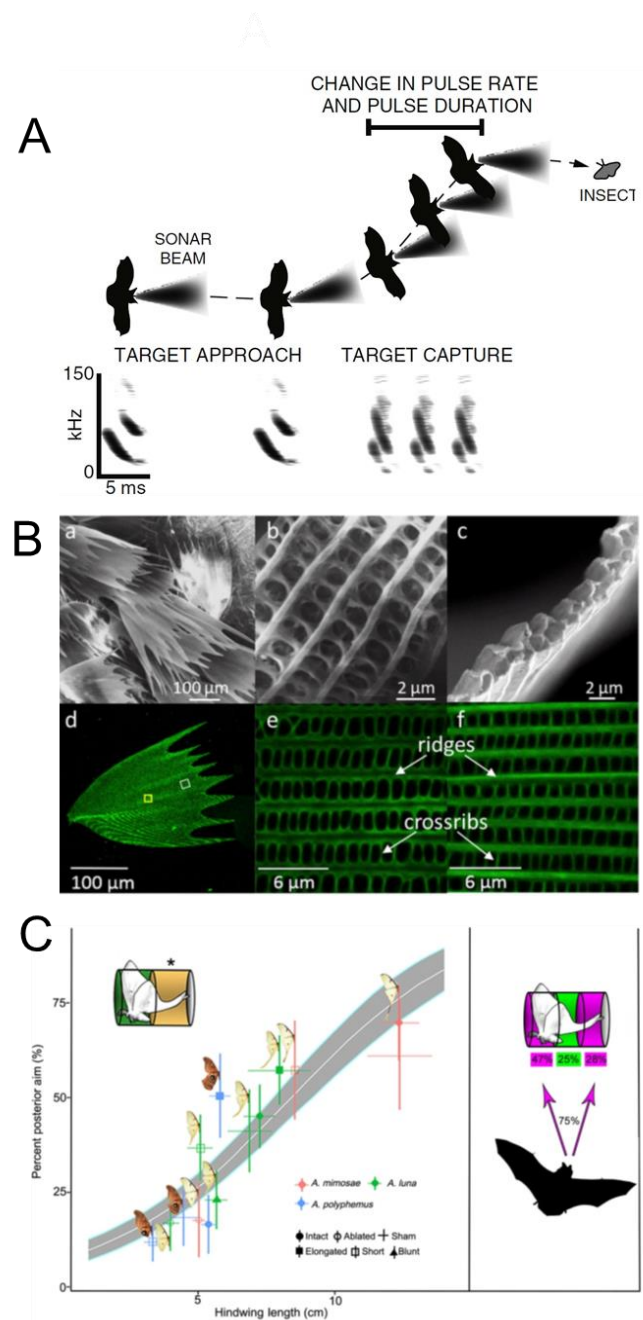

*Figure 6: Anti-predatory strategies. A) The high-resolution 3D acoustic imaging system evolved by the echolocating bats (adapted from [182]) and the moth's strategies to avoid being detected: B) appropriate scale arrangement and structure (adapted from[176]). and C) hindwing tails. Behavioural analyses reveal that (A) bats aim an increasing proportion of their attacks at the posterior half of the moth (indicated by yellow cylinder with asterisk) and that (B) bats attacked the first and third sections*

*of tailed moths 75% of the time, providing support for the multiple-target illusion. An enlarged echo illusion would likely lead bats to target the hindwing just behind the abdomen of the moth, at the perceived echo centre (highlighted in green); however, bats targeted this region only 25% of the time (adapted from [170]).*

As previously mentioned, the second type of passive acoustic camouflage developed by earless moths consists in having much of the thorax covered by hair-like scales (Figure 7A) acting as a stealth coating against bat biosonar[183–185]. As suggested by[185,186] such thoracic scales create a dense layer of elongated piliform elements, resembling the lightweight fibrous materials used in engineering as sound insulators. Their potential as ultrasound absorbers was explored in [185] by means of tomographic echo images and an average of 67% absorption of the impinging ultrasound energy emerged. Also, to provide a more in-depth investigation, the authors employed acoustic tomography to quantify the echo strength of diurnal butterflies that are, contrary to moths, not a target for bat predation. The results were then used to establish a comparison with those derived for moths (Figure 7B). Interestingly, the analysis revealed that the absorptive performance is highly influenced by the scale thickness and density, with the very thin and less dense scales typical of butterflies that can absorb just a maximum of 20% of the impinging sound energy. Conversely, the denser and thicker moth's thorax scales possess ideal thickness values that allow the absorption of large amounts of bat ultrasonic calls. These findings are confirmed by [183] where an extended list of references is also provided.

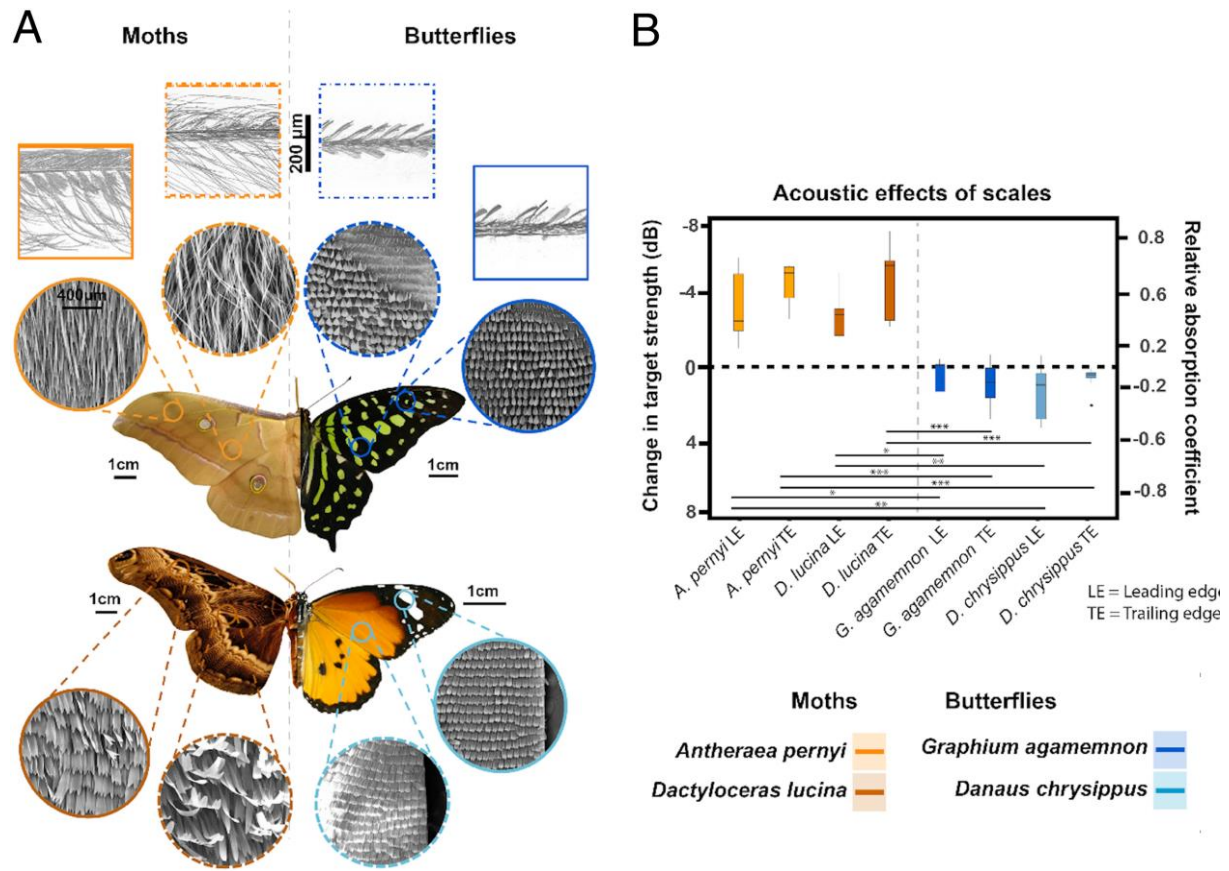

*Figure 7: Tiling patterns and acoustic effects of lepidopteran scales: A) SEM images of butterflies Graphium agamemnon and Danaus chrysippus and moths Dactyloceras lucina and Antheraea pernyi, B) Change in target strength caused by presence of scales, and equivalent intensity absorption coefficient (Adapted from[179]).*

Finally, airborne sound and vibration signals play an important role in bee communication and defence mechanisms [187]. The thorax of the bee contains a powerful musculature that is used to fly but also to produce vibratory impulses. For a long time, communication between bees seemed to be almost exclusively regulated by chemical signals, i.e. pheromones. In recent decades, it has become increasingly clear that bees live and interact in a world of sound and vibration [187,188]. One particular species, the Japanese bee *Vespa mandarinia japonica*, uses sounds to coordinate and attack predators *en masse*. In particular, the defence mechanism developed by *Vespa mandarinia* relies on the control of dorso-ventral and longitudinal muscles that do not contract alternately, as in flight, but tense simultaneously while the bee remains motionless. After a few minutes, the temperature of its thorax increases and can reach 43° (maximum temperature). If a foraging hornet tries to enter the hive, more than 500 workers quickly engulf it in a ball to rapidly raise the temperature to 47°C, which is lethal for the hornet but not to the bees [188]. This behaviour is also associated with high neural activity, underlying the bees' computation for the use and production of sounds and vibrations [189].

### 3.6 Conclusions on structures for sensing and predation

We have seen that structures that perform sensory functions are generally related with localization, allowing a certain species to either perceive its surroundings, localize prey, or escape from predators. Some common and recurring features can be found. In all cases, the sensory capability of an organism benefits from specialized transducers used to detect vibrations (e.g., cuticles for insects and arachnids, silk for spiders). Interestingly, these transducers are often associated with nonlinear constitutive behaviour, e.g., both cuticle [190] and silk [191] present a high stiffening behaviour with an exponential constitutive law. Moreover, this relationship is strongly mediated by water content, which influences the properties of both the cuticle [140,192] and silk [193]. Thus, natural structures often present a strong relationship with a fluid or viscous medium, as an agent capable to confer specific mechanical properties. Generally speaking, the sensing capability is also strongly mediated by the interaction with the substrate (e.g., trees, and leaves for spiders; sand and rocks for scorpions). Another common feature is that the interaction with the environment is also often mediated by air flows sensors, with a common hair-like shape that is present in spiders [136], scorpions [194], crickets [195], and fish [196].

### 4. Sound/vibration control, focusing and amplification

### 4.1    Echolocation in Odontocetes

Apart from communication purposes, toothed whales and dolphins (Odontocetes) use clicks, sounds and ultrasounds for sensing the surrounding environment, navigating, and locating prey [197]. This process is similar to that adopted by terrestrial animals like bats and is called echolocation [198–200]. The sounds are generated in special air cavities or sinuses in the head, can be emitted in a directional manner [201,202], and their reflections from objects are received through the lower jaw and directed to the middle ear of the animal (Figure 8)[203,204]. A number of studies have adopted CT scans and FEM to simulate sound generation and propagation in the head of dolphins or whales, demonstrating how convergent sound beams can be generated and used to direct sound energy in a controlled manner, and also how sound reception can be directed through the lower jaw to the hearing organs [205,206]. Dible et al. have even suggested that the teeth in the lower jaw can act as a periodic array of scattering elements generating angular dependent band gaps that can enhance the directional performance of the sensing process [207]. The emitted frequencies of the sounds used for echolocation are typically in the kHz range, e.g., bottlenose dolphins can produce directional, broadband clicks lasting less than a millisecond, centred between 40 to 130 kHz. Some studies have suggested that high intensity focussed sounds can even be used to disorient prey, although this remains to be confirmed [208,209]. The process of echolocation is extremely sensitive [210,211] and can provide odontocetes with a "3D view" of their surrounding environment world. This is confirmed by the fact that sonar signals employed by military vessels can confuse and distress whales and dolphins, and even lead to mass strandings [212]. Reinwald and coworkers [213] envisaged that the capability, which is still poorly-understood, of dolphins to accurately locate targets over the whole solid angle might be due to the correspondence between the reverberated coda of the signal transmitted along the bone to the ear and the location of the target that generated the signal.

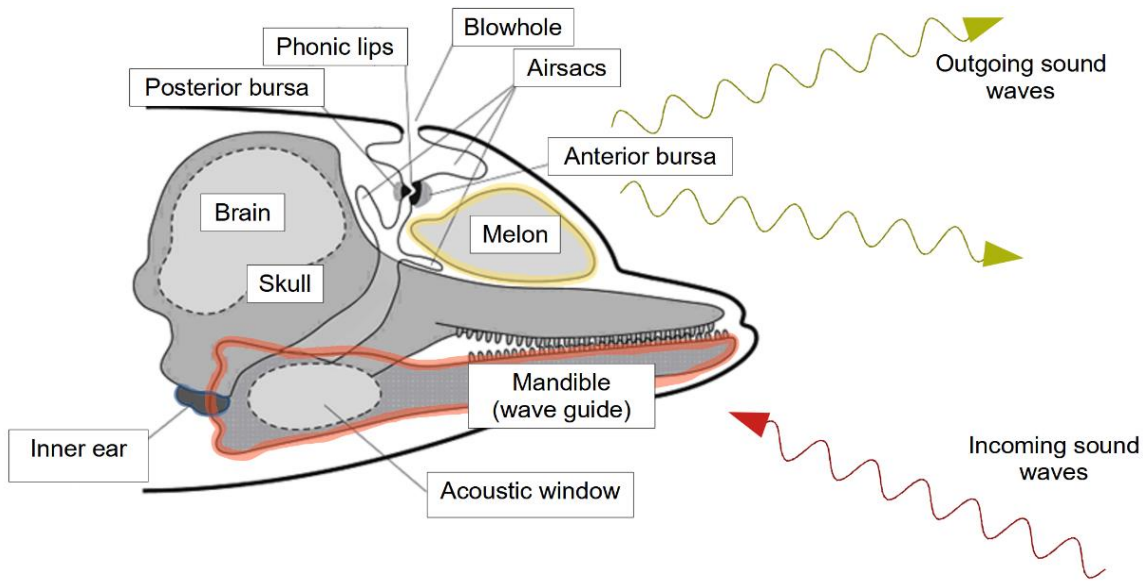

*Figure 8: Structures for sound production and detection in Dolphins. Modified and adapted from*[214].

## 4.2 High amplitude sound generation in mammals

An interesting mechanism exploited in Nature to produce sounds is to develop specific resonating structures attached to the sound-producing organs of animals with the role to selectively filter out some frequencies and amplify others [215]. There are several examples of anatomical adaptations to increase sound radiation efficiency, such as air sacs in frogs [216], birds [217], and mammals (Riede et al., 2008), or enlarged larynges in howler monkeys [218] and hammerhead bats [219]. Some animals even change their environment by constructing horns or baffles that aid in radiating the sound [220]. The case of howler monkeys (*Alouatta*) is particularly interesting: these are widely considered to be the loudest land animals, since their vocalizations can be heard clearly at a distance of 4.8 km. They emit sound at a sound level of 88dB, which means 11 dB per kg - almost 10.000 times louder per unit mass compared to other animals (Figure 9A). The function of howling is thought to relate to intergroup spacing, territory protection and social behaviour, as well as possibly mate-guarding[221]. The extraordinary capability of these monkeys to produce low frequencies and loud vocalizations has been largely studied and the exact mechanism exploited is still debated [222]. However, two main elements are considered essential in this mechanism: expansion of the hyoid bone into a large shell-like organ in the throat and large hollow air sacs located on either side of the bone (Figure 9B). When the glottis produces low frequency sounds, the hyoid and air sacs function as resonators and the constrictions in the post-glottal structures (a narrow and curved supraglottal vocal tract) reduce the velocity of the air flow, elevating its pressure and, consequently, raising its volume [223]. The harshness of the roars is a result of the forced passage of air, resulting in irregular noisy vibrations. The acoustic function of the air sacs, however, is unclear and not all authors agree on their function as resonators, proposing as an alternative an impedance matching purpose [224] or potentially to suppress resonances [225].

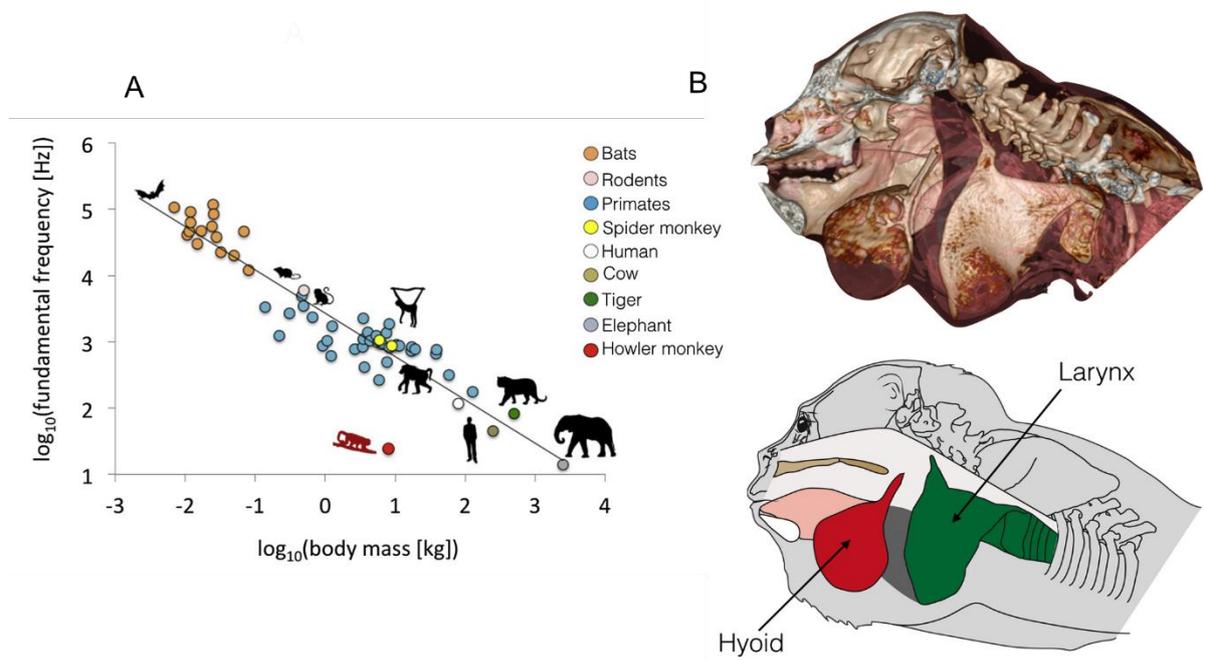

*Figure 9: A) The exceptionally low frequency of Howler Monkey vocalizations; B) Howler Monkey vocal apparatus. Adapted from [218].*

### 4.3 Cochlea in mammals

The hearing organ in mammals has developed extraordinary capabilities from the point of view of the extension of audible frequencies and perceived intensities. The human ear (Figure 10A-C), for example, is sensitive to 8 octaves of frequencies (20Hz-20kHz) and is capable of distinguishing sounds within 12 orders of magnitude of intensity (120 dB). The evolutionary complexity of this organ has represented an obstacle to the deep understanding of all the mechanisms involved and, even today, some aspects remain unexplained (for a review on the mechanical mechanisms involved see [226,227]). The *cochlea* (Figure 10E) is the core organ of the inner ear (in blue in Figure 10A), coiled in the form of a snail (hence its name) and enclosed by a bony shell. The cochlea is composed of two ducts (scala vestibuli (SV) and *scala tympani* (ST), see Figure 10B) filled with a liquid (perilymph) which is compressed by a membrane, hit by three miniscule bones of the middle ear (in red in Figure 10A). The pressure difference between the two ducts put in vibration the basilar membrane, which separates them, and which conducts a largely independent traveling wave for each frequency component of the input (this mechanism was proposed for the first time in [228] and then largely developed). Because the basilar membrane is graded in mass and stiffness along its length [229], however, each traveling wave grows in magnitude and decreases in wavelength until it peaks at a specific frequency-dependent position (see Figure 10F), thus allowing a spatial coding of the frequency contents. This is referred to as the tonotopic organization of the cochlea [230]. The mechanical vibration of the basilar membrane is then collected and translated into an electrical impulse from the hair cells (see Figure 10D) and sent to the brain for the signal decoding.

One of the most relevant and studied characteristics of the basilar membrane is that its response to an external stimulus is highly nonlinear (i.e., not proportional to the input amplitude) and this nonlinear response is also frequency specific. Moreover, each point of the cochlea has a different nonlinear response depending on the characteristic frequency pertaining to this specific point [231,232]. These features are especially evident in *in vivo* measurements, also underling the existence of an active mechanism (otoacoustic emission) added to the merely mechanical ones (see e.g., [233–235]).

The mechanisms at play are complex and often more than a possible explanation can be found in literature, but different simplified models have tried to capture the basic features of the cochlea and reproduce its incredible capacity of sensing, its tonotopic and amplification behaviours (for a review see e.g., [236,237]). One of the aspects that can be relevant for bioinspired applications in the propagation of elastic waves in solids, is the influence of the geometry (spiral) on the frequency attenuation/loss and on the tonotopic property of the sample, as also pointed out by some works (see [238,239]).

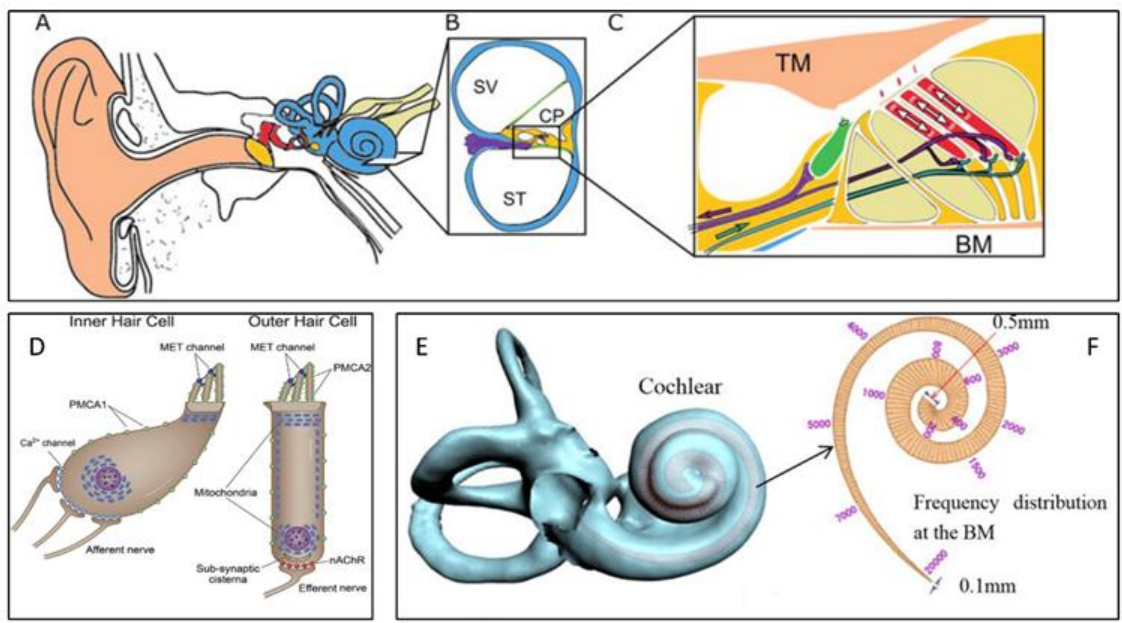

*Figure 10 : Cochlea structure. A) the outer (beige), middle (red) and inner (blue) parts of the human ear. B) Cross-section of the cochlea showing the scala vestibuli (SV) and the scala tympani (ST), separated by the cochlear partition (CP) which contains the basilar membrane (BM) and the sensory hair cells (adapted from [240]). These cells are represented in panel C in green (inner hair cells) and red (outer hair cells) and are also reported with more details in*

"Optimized structures for vibration attenuation and sound control in Nature: a review"    p. 36*subplot D (adapted from [241]). In panel E a 3D representation of the cochlea is reported and a schematic map of the tonotopic property of the basilar membrane reported in panel F (adapted from [242]).*

All these features attracted the interest of researchers working on mechanical and elastic waves manipulation devices, e.g., in the field of structural health monitoring, sensor development, guided waves, etc. There are specific works in the literature that explicitly refer to the cochlea as a bio-inspiration for metamaterial realizations and that propose acoustic rainbow sensors, where the aim is to separate different frequency components into different physical locations along the sensor (see Figure 11 and Refs. [240,242–244]). In particular, the tonotopy and the low amplitude amplifier is reproduced with a set of subwavelength active acoustic graded resonators, coupled to a main propagating waveguide in [240]. Similarly, based on a set of Helmholtz resonators arranged at sub-wavelength intervals along a cochlear-inspired spiral tube in [243], the authors realize an acoustic rainbow trapper, that exploits the frequency selective property of the structure to filter mechanical waves spectrally and spatially to reduce noise and interference in receivers. The tonotopy can be also obtained in a 3D model of the cochlea [242] by grading the mechanical parameters of an helicoidal membrane: in this case the overall cochlear is a local resonant system with the negative dynamic effective mass and stiffness.

Some of the examples of cochlea-inspiration for the design of metamaterials are shown in Figure 11. In particular, in panels A, B, C a gradient-index metamaterial for airborne sounds, made from 38 quarter-wavelength acoustic resonators of different heights is reproduced (from [240]). In panel D a rainbow trapper based on a set of Helmholtz resonators is described (from [243]). In panel E a modal analysis of a helix model of cochlea is reported, showing the different responses to different frequency excitations (in particular, at the top circle, the minimum natural frequency is 89.3 Hz; (c) at the medial circle is 5000.5 Hz; and at the base circle is 10097.2 Hz).

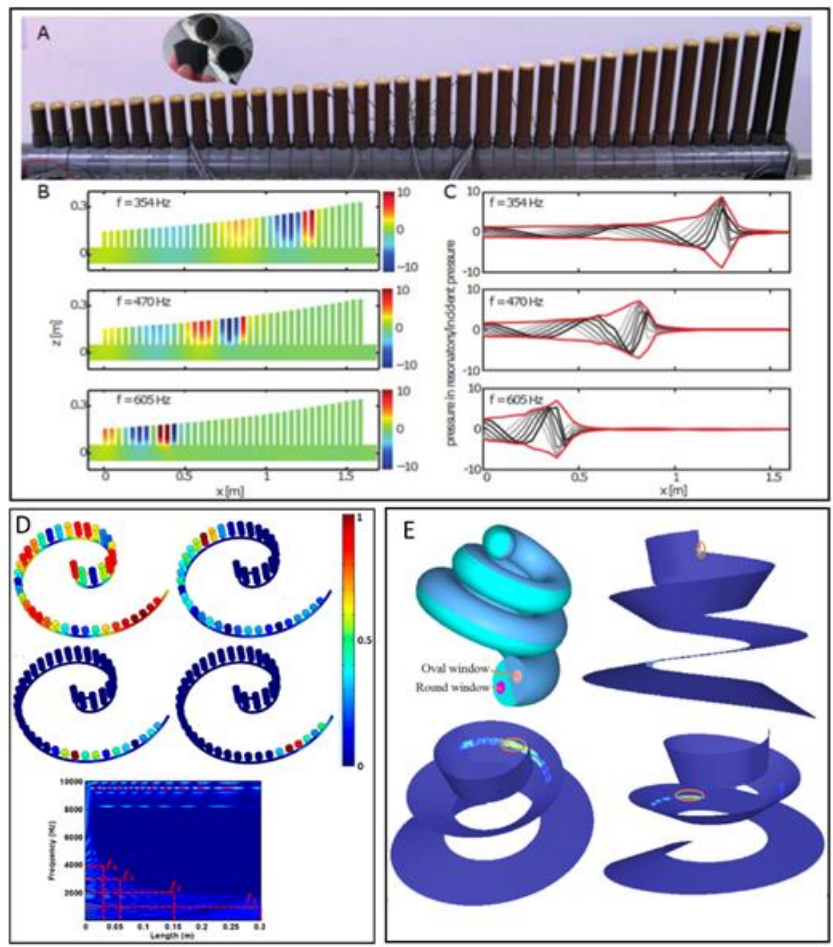

*Figure 11 : Metamaterial inspired by the cochlea. Some of the examples of cochlea-inspiration for the design of metamaterials. In particular, in panels A,B,C, a gradient-index metamaterial for airborne sounds, made from 38 quarter-wavelength acoustic resonators of different heights is reproduced (adapted from [240]). In panel D, a rainbow trapper based on a set of Helmholtz resonators is described (adapted from [243]). In panel E, a modal analysis of a helix model of cochlea is reported, showing a different response to different frequency excitations.*

**5. Natural structural evolution vs. optimization through artificial algorithms**

The structures discussed in this review are the result of optimization processes due to natural evolution spanning millions of years. Their common features are summarized in Sections 2.8 and 3.6. These evolutionary processes have however been constrained by the availability of material resources and their fabrication conditions. At the other end of the scale, there are fast developing computational algorithms used in current technology that can be used to optimize artificial materials (often bioinspired) for similar goals, where these boundary conditions can be relaxed or eliminated altogether. In artificial materials, the possibilities of design with different material combinations and distributions are virtually unbounded, and numerical algorithms can be used to optimize specific properties of nature-based architected structures [245]. The use of optimization techniques for the design of periodic structures able to attenuate vibrations, for instance phononic crystals, aims to systematically achieve objectives such as maximizing absolute band gap widths [246], normalized band gap width with respect to their central frequency [247,248], or maximized attenuation per unit length [249]. For each given combination of materials, the objective function must be evaluated through the computation of the band structure of a given unit cell configuration, using various numerical methods [250–252]. A wide variety of optimization techniques to pursue the chosen objective is available in the literature. Among these, topology optimization is one of the most employed and well-developed [253], in combination with algorithms such as Bidirectional Evolutionary Structural Optimization [254]. Another common approach is the use of genetic algorithms, an optimization scheme which is a type of evolutionary algorithm [255] and is well suited for the design of phononic crystals [256]. Another possibility is the use of machine learning tools to design structures which present desirable characteristics, i.e., using an inverse approach [257]. Many of these approaches are being more and more applied in the field of phononics [258]. The types of optimized structures emerging from these algorithms have some common traits with naturally evolved structures, and some distinctive differences. On the one hand, recurring features are many of those cited in Section 2.8: heterogeneity, porosity, hierarchical organization, efficient resonating structure, graded properties, in some cases chirality [259,260]. In this case, artificial optimization techniques can improve existing bioinspired design for specific objectives. On the other hand, implementing unconstrained numerical optimization can enable a wider exploration of the phase space, potentially leading to exotic designs with little resemblance to existing biological structures. However, this is not surprising, since optimization based on natural evolution is in most cases a muti-objective process, where different properties are simultaneously addressed (e.g., quasistatic strength/toughness and dynamic attenuation).

## 6. Conclusions

In conclusion, we have presented a review of some notable examples of biological materials exhibiting optimized non-trivial structural architectures to achieve improved vibration control or elastic wave manipulation, for many different purposes. The fields in which these features appear are mainly impact and vibration damping and control, communication, prey detection or mimesis, and sound amplification/focusing. From the documented cases, some recurrent strategies and structural designs emerge. Among them, an important feature is hierarchical structure, which appears to be essential to enable effects at multiple scale levels, and therefore in multiple frequency ranges. Moreover, these recurrent structural features appear at very different size scales (from microns to meters), in disparate environments (terrestrial or marine) and for different functions. This is an indication that the designs are particularly resilient and effective in their purposes, which encourages the adoption of a biomimetic approach to obtain the comparable types of optimized dynamic mechanical properties in artificial structures. This is a particularly attractive proposition in the field of phononic crystals and acoustic metamaterials, which have recently emerged as innovative solutions for wave manipulation and control, and where a biomimetic approach to design has thus far been limited to a few cases, especially considering that biological materials derive from self-assembly, so that are inherently periodic or hierarchical in structure. In general, further investigations in the natural world will no doubt continue to reveal original architectures, designs, and advanced functionalities to be exploited for metamaterials and other vibration-control technologies, where function (or multiple functions) is/are achieved through form and structure.

**Acknowledgments**

All authors are supported by the European Commission H2020 FET Open "Boheme" grant no. 863179.

**Author contributions**

Conceptualization, F.B and N.M.P. All authors contributed to the writing of the manuscript.

**Declaration of interests**

The authors declare no competing interests.

**Figure titles and legends**

*Figure 1: Mantis shrimp*

a) Peacock mantis shrimp, with highlighted raptorial dactyl clubs to strike hard objects (adapted from [18]); b) Morphological features of the clubs, in cross-section view, divided in an impact region, a periodic region and a striated region; c) Scanning electron micrograph of the coronal cross-section, showing reinforcing fibre helicoidal arrangement; d) schematic of a Finite Element Analysis model accounting for graded material properties (adapted from [19]).

*Figure 2: Vibration attenuation in the woodpecker skull (adapted from [31])*

(A): Volume fraction ratio of skull bone, local measured modulus, and macro-equivalent modulus around the skull. (B): 3D finite-element model of the skull and hyoid bone. Note that the Young's modulus on the skull is not uniform. (C): first ten modes of the skull under a pre-tension on the hyoid in the range 0-25 N. (D), upper panel: stress wave at a brain location under impact direction. (d), lower panel: stress spectrum in the frequency domain obtained by FFT.

*Figure 3: Modelling of vibration attenuation in the woodpecker skull (adapted from [10])*

(A): lumped-elements model of the head of a woodpecker. (B): empirical model of the spongy bone by means of an aluminium enclosure filled with glass microspheres. (C): vibration transmissibility as a function of frequency for different diameters of the $SiO_2$ microspheres.

*Figure 4: Biological systems with suture tessellation*

(a) examples from Flora and Fauna (adapted from [58]); (b) the leatherback turtle shell (adapted from [60])

*Figure 5: Prey sensing similarities in spiders and scorpions*

*a) web structure: a typical orb web of a spider Nuctenea umbratica. The web is built by means of junctions between threads and surfaces, b) junctions between radial threads, c) and junctions between radial and spiral threads. A flying prey can be eventually detected by air flow sensors, d) the tricobothria. If the prey impacts the web, the vibrational signal will be transmitted mainly by radial threads and be perceived by e) lyriform organs of the spider. Figure adapted from [93,132]. f) schematic of scorpion prey detection using surface waves; g) sensory hairs and mechanoreceptors located at the slit sensilla sense surface waves. Adapted from [162159]*

*Figure 6: Anti-predatory strategies*

*(a) The high-resolution 3D acoustic imaging system evolved by the echolocating bats (adapted from [174]) and the moth's strategies to avoid being detected: (b) appropriate scale arrangement and structure (adapted from [178]). and (c) hindwing tails. Behavioural analyses reveal that (A) bats aim an increasing proportion of their attacks at the posterior half of the moth (indicated by yellow cylinder with asterisk) and that (B) bats attacked the first and third sections of tailed moths 75% of the time, providing support for the multiple-target illusion. An enlarged echo illusion would likely lead bats to target the hindwing just behind the abdomen of the moth, at the perceived echo center (highlighted in green); however, bats targeted this region only 25% of the time (adapted from ([261]).*

*Figure 7: Tiling patterns and acoustic effects of lepidopteran scales*

*(a) SEM images of butterflies Graphium agamemnon and Danaus chrysippus and moths Dactyloceras lucina and Antheraea pernyi, (b) Change in target strength caused by presence of scales, and equivalent intensity absorption coefficient, (c) Change in target strength caused*

*by presence of scales, and equivalent absorption coefficient as a function of wing thickness/wavelength. (Adapted from[185])*

**Figure 8: Sound generation in Howler monkeys**

*a) The exceptionally low frequency of Howler Monkey vocalizations; b) Howler Monkey vocal apparatus. Adapted from [218].*

**Figure 9 : Metamaterial inspired by the cochlea**

*Some of the examples of cochlea-inspiration for the design of metamaterials. In particular, in panels A,B,C, a gradient-index metamaterial for airborne sounds, made from 38 quarter-wavelength acoustic resonators of different heights is reproduced (adapted from [240]). In panel D, a rainbow trapper based on a set of Helmholtz resonators is described (adapted from [243]). In panel E, a modal analysis of a helix model of cochlea is reported, showing a different response to different frequency excitations.*